\documentclass[amsmath,amssymb]{revtex4}
\usepackage{graphicx}
\usepackage{dcolumn}
\usepackage{bm}

\usepackage{makeidx}
\usepackage{epsfig}
\usepackage{color}
\usepackage{amsmath}
\makeindex
\begin{document}

\title{{Modelling of inelastic effects in molecular electronics}\footnote{Based on a talk presented at the
conference ``Progress in Nonequilibrium Green Functions III, Kiel,
Germany, 22. -- 25. August 2005''}}

\author{Antti-Pekka Jauho}

\address{NanoDTU, Department of Micro and Nanotechnology \\
Technical University of Denmark, Bldg. 345East, {\O}rsteds Plads, \\
DK-2800 Kgs. Lyngby, Denmark\\
E-mail: antti@mic.dtu.dk}


\begin{abstract}
Ab initio modeling of molecular electronics is nowadays routinely
performed by combining the Density Functional Theory (DFT) and
Nonequilibrium Green function (NEGF) techniques. This method has
its roots in the current formula given by Meir and Wingreen, and
we discuss some applications and accompanying pitfalls and
restrictions of this approach. Quite recently papers have begun to
appear where inelastic effects are considered, and we illustrate
these new developments by describing our own work on transport in
atomic gold wires.
\end{abstract}

\maketitle

\section{Introduction}\label{sec:intro}
\subsection{Some background, and a few historical remarks}\label{subsec:background}

In my previous two reviews in this conference
series\cite{Jauh00,Jauh03} I have discussed applications of a
certain current formula, due to Meir and Wingreen\cite{Meir92}, to
mesoscopic structures.  The present review will continue along the
same lines.  As was pointed out in my review in 2002, this formula
``is gaining widespread use"; this has indeed been the case also
in the intervening three years: more than 350 papers have appeared
applying this formula or its variants, and an exhaustive review is
clearly out of the question.  I have therefore restricted myself
to two topics which I  believe to be very important, not only in
pure academic sense but also as far as applications are concerned.
Specifically, I will describe {\it ab initio} modeling of
molecular electronics, where one combines the density functional
theory (DFT) with the nonequilibrium Green function technique
(NEGF), and, secondly, I address the difficulties, challenges, and
successes in the very recent efforts where inelastic effects have
been included in the modeling.

\subsection{The basic equations, and their limitations}

The details of the derivation of the Meir-Wingreen formula have
been given several times before\cite{Jauh94,Haug96}, and there is
no need to repeat them here, and we just summarize the basic
philosophy.  In the spirit of the scattering theory approach to
conductance\cite{Land57,Land70,Butt86}, one considers a system
(say, a molecule, or a quantum dot) which is coupled via ideal
leads to reservoirs, which are so large that they can be described
by an equilibrium distribution.  The nonequilibrium aspects are
introduced via the following construction, pioneered by Caroli et
al.\cite{Caro71a,Caro71b,Caro72,Comb71}: in the infinite past the
various subsystems are separated, with their respective chemical
potentials, and the couplings between the various subsystems are
turned on adiabatically.  The couplings are treated
perturbatively, but to all orders.  The double contour of Keldysh
enters because only the state in the remote past is known.  No
assumptions about the smallness of the couplings are needed.  One
may have to perform a self-consistent calculation of the various
parameters describing the system.  Below we have some comments
concerning this construction.

Let us next consider some generic Hamiltonians:
$H=H_L+H_R+H_T+H_{\rm cen}$, or, explicitly:
\begin{eqnarray}
H &=& \sum_{k,\alpha\in L/R} \epsilon_{k,\alpha}
c^\dagger_{k,\alpha} c^{\phantom\dagger}_{k,\alpha}\nonumber\\
&\quad& + \sum_{k,\alpha\in L/R;n} \left[V_{k\alpha;n}
c^\dagger_{k,\alpha} d^{\phantom\dagger}_n + {\rm h.c.}\right] +
H_{\rm cen}\left[\{d^{\phantom\dagger}_n\},
\{d^\dagger_n\}\right]\;,
\end{eqnarray}
where the central part Hamiltonian must be chosen according to the
system under consideration.  The operators
$\{d_n\},\{d^\dagger_n\}$ refer to a complete set of
single-particle states of the central region.  The derivation of
the basic formula for the current does not require an explicit
form for $H_{\rm cen}$; the actual evaluation of the formula of
course requires this information.  We write $H_{\rm cen}=\sum_n
\epsilon_n d^\dagger_n d^{\phantom\dagger}_n + H_{\rm int}$, where
$H_{\rm int}$ could be electron-phonon interaction,
\begin{equation}
H_{\rm int}^{\rm el-ph} = \sum_{m\sigma} d^{\dagger}_{m,\sigma}
d^{\phantom\dagger}_{m,\sigma} \sum _{\bf q} M_{m,\bf q}
\left[a^\dagger_{\bf q} + a^{\phantom\dagger}_{\bf
q}\right]\;,\label{Helph}
\end{equation}
or an Anderson impurity:
\begin{equation}
H_{\rm int}^A = U \sum_m d_{m,\uparrow}^\dagger
d^{\phantom\dagger}_{m,\uparrow} d_{m,\downarrow}^\dagger
d^{\phantom\dagger}_{m,\downarrow} \label{HAnd}\;,
\end{equation}
or, perhaps, some other model for the Coulomb interaction.

The current operator for the (say) left lead is
\begin{equation}
I_L=-\frac{ie}{\hbar}\sum_{k,n}\left[-V_{kL;n}c^\dagger_{kL}d^{\phantom\dagger}_n+
V^*_{kL;n}d^\dagger_nc^{\phantom\dagger}_{kL}\right]\;.\label{currop}
\end{equation}
Many physically relevant observables can be expressed in terms of
expectation values of the current operator, or its higher powers.
For example, one can show\cite{Jauh94,Haug96} that the current
leaving the left contact is
\begin{eqnarray}
\langle I_L \rangle=J_L(t) &=& - {2e\over \hbar} \int_{-\infty}^t
dt_1 \int {d\epsilon\over 2\pi} {\rm ImTr} \Big \{
e^{-i\epsilon(t_1-t)} {\bf \Gamma}^L(\epsilon,t_1,t)
\nonumber\\
&\quad&\quad\times \left[ {\bf G}^<(t,t_1) + f^0_L(\epsilon) {\bf
G}^r(t,t_1)\right] \Big \}\;. \label{jtime}
\end{eqnarray}
Here the Green functions are defined by
\begin{eqnarray}
G_{nm}^<(t,t_1)&=&i\langle d_m^\dagger(t_1) d^{\phantom\dagger}_n(t)\rangle\label{Gcorr}\\
G_{nm}^r(t,t_1)&=&-i\theta(t-t_1)\langle
[d^{\phantom\dagger}_n(t),d_m^\dagger(t_1)]\rangle\label{Gret}\;,
\end{eqnarray}
$\Gamma_{mn}$ describes the coupling between the central region
and the contacts, and $f_L^0(\epsilon)$ is the equilibrium
distribution function of the left contact.  The trace in
Eq.(\ref{jtime}) is over the elements of the product of the
$\mathbf\Gamma$- and $\mathbf G$-matrices.  Further, in
Eq.(\ref{jtime}) we allow for time-dependent couplings or
single-particle energies. Thus, Eq.(\ref{jtime}) is applicable for
charge pumping, which has received a lot of recent attention (see,
e.g., Aono\cite{aono04} or Kashcheyevs et al\cite{kash04}). In the
dc-limit the time-dependence drops out from the
$\mathbf\Gamma$-matrix, and the time-integral generates the
energy-representations of the $\mathbf G$-matrices, which now
depend only on the time-difference.  After symmetrization (in
dc-limit the current through left and right barriers are equal; in
our notation $J_R=-J_L$), Eq.(\ref{jtime}) then reduces to the
result of Meir and Wingreen\cite{Meir92}:
\begin{eqnarray}
J &=& {ie\over 2\hbar} \int {d\epsilon\over
2\pi} {\rm Tr}\Big \{ \left[ {\bf \Gamma}^L(\epsilon) - {\bf
\Gamma}^R(\epsilon)\right]
{\bf G}^<(\epsilon)\nonumber\\
&\quad&\quad + \left[ f_L^0(\epsilon){\bf \Gamma}^L(\epsilon) -
f_R^0(\epsilon) {\bf \Gamma}^R(\epsilon)\right] \left[ {\bf
G}^r(\epsilon) - {\bf G}^a(\epsilon)\right]\Big\}\;. \label{jprop}
\end{eqnarray}

The expressions (\ref{jtime}) and (\ref{jprop}) are the central
formal results whose consequences we explore in this review. They
are formally exact, and give the {\it tunneling} current for an
interacting system coupled to noninteracting contacts (or, more
precisely, for contacts which can be described by an effective
single-body Hamiltonian).  The total current flowing in the
outside circuit may contain contributions from displacement
currents\index{Displacement current}, and these must be considered
separately, see, e.g., Anantram and Datta\cite{anan95}, or Wang et
al.\cite{wang99} . The physical interpretation of (\ref{jprop})
can be made more transparent by writing it in an alternative form
(consider here only current through the left barrier):
\begin{equation}
J_L=\frac{e}{h} \int d\epsilon {\rm Tr}\left\{{\bf
\Sigma}^{L,<}(\epsilon){\bf G}^>(\epsilon)-{\bf
\Sigma}^{L,>}(\epsilon){\bf G}^<(\epsilon)\right\},
\label{jThomas}
\end{equation}
where the left self-energy gives the coupling to the left lead.
The first term in (\ref{jThomas}) gives the current flowing
through the left barrier towards the central region (because it is
proportional to $G^>$, i.e., the empty states in the central
region), while the second term gives the current flowing through
the left barrier towards the left contact (because it is
proportional to $G^<$, i.e., the occupied states in the central
region.  Likewise, the self-energies are proportional to the
occupied lead states, and the empty lead states, respectively.

It should also be noted that these equations only {\it define} the
starting point of any calculation: to get physical results one
must evaluate the correlation function and the retarded/advanced
Green function, Eqs.(\ref{Gcorr}) and (\ref{Gret}), respectively.
These functions obey the Keldysh equation,\index{Keldysh equation}
and the (nonequilibrium) Dyson equation:\index{Dyson equation}
\begin{eqnarray}
G^<&=&(1+G^r\Sigma^r)G^<_0(1+\Sigma^aG^a)+G^r\Sigma^<G^a,\label{Keldysh}\\
G^r&=&G^r_0+G^r_0\Sigma^rG^r\label{Dyson}.
\end{eqnarray}
The first term in the equation for $G^<$ represents a boundary
term, which is often omitted in discussions of a stationary
situation, since it only affects the transient behavior.  The
success of the theory depends on whether one can construct a
self-energy functional\index{Self-energy functional} that captures
the essential physics, and that a good solution can be found for
these coupled equations. Both of these steps may be hard indeed.

While the above approach is widely used, one should not forget its
inherent limitations.  Consider first the idea of partitioning the
system into noninteracting leads, and an interacting central
region. This separation is crucial for obtaining the compact
result for the current.  But what are the physical criteria for
selecting the boundaries between the three regions?  In our first
{\it ab initio} calculations of inelastic effects in atomic gold
wires\cite{Fred04}, to be discussed below, we allowed only the
four atoms constituting the wire to move around their equilibrium
positions and thereby define the interacting central region, while
in later work\cite{Fred05} several additional atoms from the
contact region were also taken to be a part of the central region.
Including the atoms from the contacts changes the results only
slightly, but in our case it was the computational effort that
dictated how many contact atoms were included, not an {\it a
priori} physical criterion.  In a similar vein, allowing the
Coulomb interactions to be present only in the central region
appears somewhat arbitrary.  Screening will of course diminish the
strength of the Coulomb interactions once the contact has
broadened into the bulk-like electrode, but once again it is not
clear how large the interacting region should be. There are also
certain subtleties related to overall charge neutrality of the
system in nonequilibrium conditions, see Mera et. al \cite{Mera05}
for some illustrative model calculations.

The assumption of having noninteracting leads at their respective
chemical potentials also hides some important physics.  The hot
charge carriers injected into the collecting electrode are assumed
to dissipate their energy, without heating up the contacts.  There
are no terms in the Hamiltonian to describe energy relaxation in
the contacts, nor are there terms which describe the coupling of
the contacts to the surrounding thermal bath.

The idea of having the system separated into independent parts in
infinite past is important in the formal development of the
theory: one needs a well-defined initial state which is then
modified by external perturbations.  In an experiment one does not
usually introduce the couplings in an adiabatic way at a fixed
external bias, and watch the build-up of the current flow until it
settles to a steady state, rather the entire system is in
equilibrium until the driving voltage is coupled to the system.
This has been recently recognized by Stefanucci and
co-authors\cite{Stef04,Kurt05}, who develop an approach based on
time-dependent density functional theory, which avoids the
partitioning.  These results will be discussed elsewhere in these
proceedings.

\section{Uses and misuses of the Meir-Wingreen formula}

\subsection{Mean-field theories}
The calculation of the current using (\ref{jprop}) together with
the noneqilibrium Keldysh and Dyson equations (\ref{Keldysh}) and
(\ref{Dyson}), respectively, is a very difficult task for any
nontrivial interaction (hidden in the self-energies), and it is
natural to look for physically motivated simplifications. Let us
first consider the case where the interactions are of the
mean-field type, such as e.g. in the density-functional approach.
Mean-field self-energies are one-point functions, with vanishing
Keldysh components, and the Keldysh equation becomes a formally
explicit expression for ${\mathbf G}^<$,
\begin{equation}
{\mathbf G}^<={\mathbf G}^r i[{\mathbf\Gamma}^L f^0_L +
{\mathbf\Gamma}^R f^0_R]{\mathbf G}^a.\label{mean}
\end{equation}
Note however that the ${\mathbf G}^{r,a}$ may contain some
functional of ${\mathbf G}^<$ as a parameter (the level occupation
is a typical example, $n_n=-iG^<(t,t)_{nn}$; another example is
Eq.(\ref{SigmaHr}) below), and a self-consistent calculation may
be needed. Nevertheless, Eq.(\ref{mean}) leads to a very compact
expression for the current, as we now shall show. Using
(\ref{mean}) and the relation ${\mathbf G}^r-{\mathbf
G}^a=-{\mathbf G}^r({\mathbf \Sigma}^a -
{\mathbf\Sigma}^r){\mathbf G}^a\equiv -i{\mathbf
G}^r{\mathbf\Gamma}{\mathbf G}^a$ (here
${\mathbf\Gamma}={\mathbf\Gamma}^L+{\mathbf\Gamma}^R$), one finds
that the current formulas (\ref{jprop}) and (\ref{jThomas}) reduce
to
\begin{equation}
J_L=\frac{e}{h}\int d \epsilon T_{\rm
tot}(\epsilon)[f_L^0(\epsilon)-f_R^0(\epsilon)], \label{jnonint}
\end{equation}
where
\begin{equation}
T_{\rm tot}(\epsilon)={\rm Tr}\left\{{\bf \Gamma}^L(\epsilon){\bf
G}^r(\epsilon){\bf \Gamma}^R(\epsilon){\bf G}^a(\epsilon)\right\}
\label{transprop}
\end{equation}
is the transmission probability. Eq.(\ref{jnonint}) is used very
frequently and occasionally callously: one must bear in mind that
it only holds for a noninteracting, or a mean-field theory.
Nevertheless, it is extremely useful, and forms the basis even for
some commercial software, which is being used for first-principles
modeling of molecular electronics.

\subsection{Conservation laws}
As emphasized above, the present approach becomes useful the
moment one has constructed a self-energy functional. But not all
self-energy functionals are allowable: they must be such that
current conservation is obeyed.  This important consistency check
can be formulated as follows\cite{Thomas}.  The self-energy
appearing in the Keldysh equation contains contributions both from
the coupling to the leads, and from the interactions in the
central region.  We thus write ${\bf\Sigma}_{\rm
tot}={\bf\Sigma}_{\rm int}+\sum_{\alpha\in L/R}
{\bf\Sigma}^\alpha$. Using the Keldysh equation repeatedly, one
can show that
\begin{equation}
{\rm Tr}\left\{{\bf\Sigma}^<_{\rm tot}{\bf G}^>-{\bf\Sigma}^>_{\rm
tot}{\bf G}^<\right\}\equiv 0. \label{condtion}
\end{equation}
Using this condition, one can show that the current conservation
condition $\sum_\alpha J^\alpha =0$ leads to
\begin{equation}
\int d \epsilon {\rm Tr}\left\{{\bf\Sigma}_{\rm
int}^<(\epsilon){\bf G}^>(\epsilon)-{\bf\Sigma}_{\rm
int}^>(\epsilon){\bf G}^<(\epsilon)\right\}=0.\label{curcon}
\end{equation}
In a practical calculation the self-energy is known often only
numerically, and Eq.(\ref{curcon}) can be used to check that
numerical errors have not led to an unphysical result.  Using the
language of kinetic theory, Eq.(\ref{curcon}) says that the
integrated collision term must vanish - a condition familiar from
Boltzmann-like theories.  A typical example of a "good"
self-energy is the self-consistent Born approximation with free
equilibrium phonons (which we use below in our discussion of
inelastic effects in Au-nanowires): a simple calculation shows
that the condition (\ref{curcon}) is identically satisfied in this
case.  Calculations considering  an interacting {\it and}
nonequilibrium phonon distribution are scarce, even though initial
steps towards these goals are being taken\cite{Rynd05}.

\subsection{The wide-band limit}
In the wide-band limit, when the energy-dependence of the
$\Gamma$'s can be neglected, the Meir-Wingreen formula assumes a
particularly simple form (for simplicity we consider here a
one-state model for the central region):
\begin{equation}
J=\frac{e}{h}\frac{\Gamma^L\Gamma^R}{\Gamma^R+\Gamma^L}\int
d\epsilon[f_L(\epsilon)-f_R(\epsilon)]A(\epsilon),
\label{WBLcurrent}
\end{equation}
where $A(\epsilon)=-2{\rm Im}G^r(\epsilon)$ is the interacting
spectral function. Usually an exact evaluation of the spectral
function is not possible, and one looks for approximate methods.
An often used approach is to start from the atomic limit, i.e., an
isolated central region, for which the Green function can be
calculated, either analytically or perhaps by an exact
diagonalization, and then broaden the sharp levels by a
phenomenological width. Important examples include a model where a
single level is coupled to phonons, for which the Green function
is\cite{Maha90}
\begin{equation}
G^r(t)=-i\theta(t)\exp[-it(\epsilon_0-\Delta)-\Phi(t)],
\end{equation}
with
\begin{eqnarray}
\Delta&=&\sum_q \frac{M^2_q}{\omega_q}\nonumber\\
\Phi(t)&=&\sum_q
\frac{M^2_q}{\omega_q^2}[N_q(1-e^{i\omega_qt})+N_q(1-e^{-i\omega_qt})],
\end{eqnarray}
or an isolated Anderson impurity, for which one has
\begin{equation}
G^\sigma(\epsilon)=\frac{\langle
n_{\bar\sigma}\rangle}{\epsilon-\epsilon_\sigma-U}+\frac{1-\langle
n_{\bar\sigma}\rangle}{\epsilon-\epsilon_\sigma}.
\end{equation}
A tempting approach would be to add a factor $-\Gamma t/2$ in the
exponent in the phonon problem, or a factor $i\Gamma/2$ in the
denominator in the Anderson impurity problem.  This would however
lead to erroneous physics: in the phonon case the heuristic
procedure implies that the Fermi seas in the contacts has been
ignored (for an extended discussion on this issue, see
Ref.\cite{Flen03}), and in the Anderson impurity case higher
correlations  beyond Hartree-Fock level would be lost (see Sect.
12.7 in Ref.\cite{Haug96}). These simple examples are mentioned
here in order to stress that extreme care is needed in treating
the interacting problems, and that seemingly innocent steps may
have far-reaching consequences.

\section{Density functional method for nonequilibrium electron
transport}\label{sec:DFT} In this section I briefly discuss some
of the generic features of theories that combine some {\it ab
initio} electron structure theory and a nonequilibrium transport
theory.  In the literature one can find several implementations,
here I focus on the TRANSIESTA code, developed at the Technical
University of Denmark\cite{Bran02}.

Most electronic structure calculations are restricted in the sense
that the geometry must be finite, or periodic, and that the
electronic system is in equilibrium. The situation is very
different, say, in the case of molecular electronics: here a small
subsystem lacking translational invariance couples to
semi-infinite leads {\it and} the electronic subsystem can be far
from equilibrium.  Ideally one should describe the whole system
(the central region and electrodes) on equal footing.  As is
well-known, the Density Functional Theory gives the exact
electronic density and total energy, if the exact
exchange-correlation functional was known.  Since this is not the
case, one must resort to approximate forms of the functional, such
as the local-density approximation (LDA), or the generalized
gradient approximation (GGA), or something else.  There is no
theory to say which (approximate) functional is the best, rather
the choice is made based on painstaking tests, and comparisons in
some limits where alternative methods, or experiments, can give
benchmarks. In an attempt to extend DFT to nonequilibrium
situations one must go one step further: the Kohn-Sham
single-particle wave-functions are used when calculating the
current.  Thus, genuinely many-particle effects, such as the Kondo
effect, are excluded from the treatment.  Inelastic effects can be
included, as discussed in the next section.  An improvement of the
present approach could conceivably be reached by the
current-density formalism\cite{Vign96}.

At the core of the DFT-NEGF implementation described in
Ref.\cite{Bran02} is the SIESTA code\cite{Sanc99}.  This approach
has been tested in a large number of applications with excellent
results, and it has many technical advantages because of the
employed finite range orbitals for the valence electrons: not only
do the numerics get faster but also the system partitioning
becomes unambiguous.  The SIESTA approach can be extended to
nonequilibrium by using a nonequilibrium electron density as an
input; the nonequilibrium electron density is calculated with the
help of NEGF:
\begin{equation}
n(x)=-i G^<(x=x',t=t')=\int\frac{d\epsilon}{2\pi
i}G^<(x=x',\epsilon),\label{n(x)}
\end{equation}
where $G^<$ follows directly from the Keldysh equation, because
the self-energy is a known function for mean-field theories.
Explicitly, we recall that the lesser self-energy has the
structure
\begin{equation}
\Sigma^<=i(\Gamma^L f_L + \Gamma^R f_R), \label{Sigmalesser}
\end{equation}
where the occupation factors of the left and right contacts
contain the information about the potential difference. Hence, all
that one needs are the retarded and advanced Green functions, and
these are obtained by evaluating
\begin{equation}
{\bf G}^{r,a}(E)=\left[E{\bf I}\pm i\eta-{\bf H}\right]^{-1},
\label{bfG}
\end{equation}
where
\begin{equation}{\bf H}=
   \begin{pmatrix}
    {\bf H}_L+{\bf \Sigma}_L \quad\ & {\bf V}_L & 0 \\
    {\bf V}_L^\dagger & {\bf H}_C & {\bf V}_R \\
    0 & {\bf V}_R^\dagger & \quad{\bf H}_R+{\bf \Sigma}_R
    \
  \end{pmatrix},
\end{equation}
with an obvious nomenclature.  The semi-infinite left and right
leads are accounted for by the self-energies ${\bf\Sigma}_{L/R}$,
see, e.g., the book by Datta\cite{Datt95}.  Importantly, to
determine ${\bf V}_R$, ${\bf V}_L$, or ${\bf H}_C$ one does not
need to evaluate the density matrix outside the $L-C-R$ region, if
the $L-C-R$ region is defined so large that all screening takes
place inside it.

Summarizing, and somewhat simplifying, the TRANSIESTA iterative
loop consists of the steps
\begin{equation}
{\rm initial} \, n(x)\Rightarrow {\rm SIESTA} \Rightarrow
\psi_{\rm KS}(x) \Rightarrow {\rm NEGF} \Rightarrow {\rm new} \,
n(x),
\end{equation}
and the iteration is repeated until convergence is achieved for
the desired quantity, such as the current for a given voltage
difference.  For a detailed description of many of the technical
details suppressed here we refer to the paper by Brandbyge {\it et
al.}\cite{Bran02}.  The scheme outlined above has been applied by
a large group of researchers to many specific physical systems.
Occasionally the agreement with experiments reaches a quantitative
level, which is indeed very satisfying, while sometimes the
predicted current can be orders of magnitude too large.  At the
moment of writing this review, there is no consensus of whether
the discrepancies are due to poorly controlled experiments, bad
implementations of the DFT-NEGF scheme, or due to an inadequacy of
the entire concept.  A possible cause for the discrepancy has very
recently been identified by Toher et al.\cite{Tohe2005}, who
suggest that self-interaction corrections (which are not included
in the GGA-LDA underlying most theoretical work, including ours)
could remedy some of the problems. Nevertheless, a lot of research
remains to be done.

\section{Inelastic scattering and local heating in atomic gold
wires}\label{sec:inelas}The issue of vibrational effects in
molecular electronics has recently drawn a lot of interest because
inelastic scattering and energy dissipation inside atomic-scale
conductors are of paramount importance for device characteristics,
working conditions, and their
stability\cite{Kush04,Wang04,Smit04,Flen03}.  Inelastic effects
are important, not only because of their potentially detrimental
influence on device functioning, but also because they can open up
new possibilities and operating modes.  Vibrational effects are
often visible in the measured conductances of nanoscopic objects;
here we focus on recent experimental studies on free standing
atomic gold wires\cite{Agra02}.   Agra{\"i}t and co-workers used a
cryogenic STM tip to first create an atomic-scale gold wire
(lengths up to seven gold atoms have been achieved), and then
measured its conductance as a function of the displacement of tip,
and the applied voltage.  The data showed clear drops of
conductance at a certain voltage, and the interpretation was that
an excitation of an inelastic mode was taking place, leading to
enhanced back-scattering, and hence drop in the conductance.  It
should be pointed out that opening a new vibrational mode in the
atomic scale conductor does not necessarily lead to a decrease in
conductance (one can envisage various assisted processes), and a
proper theory should be able to predict conductance enhancement as
well, whenever the physics dictates so.

Here I will briefly describe our recent work\cite{Fred04} on
inelastic effects in the kind of wires studied by Agra{\"i}t et
al.\cite{Agra02}.  An important aspect of our work is that we go
beyond lowest order perturbation theory in the electron-vibration
coupling, and therefore polaronic effects can be included.  We
also address the issue of phonon heating, albeit within a
phenomenological model, as discussed below.

The calculational method consists of three steps.  (i) The
mechanical normal modes and frequencies of the gold chain are
evaluated.  (ii) The electronic structure and electron-vibration
coupling elements are evaluated in a localized atomic-orbital
basis set.  (iii) The inelastic transport is evaluated using NEGF
by using a self-consistent Born approximation self-energy in the
Dyson and Keldysh equations for the respective Green functions.
The electrical current and the power transfer are then evaluated
with (here, for the left lead; for a detailed derivation, see
Ref.\cite{Thomas})
\begin{eqnarray}
I_L &=& \frac{e}{h}\int d\epsilon\, t_L (\epsilon)\\
P_L &=& \int\frac{d\epsilon}{2\pi\hbar}\,\epsilon t_L(\epsilon)\label{PL}\\
t_L(\epsilon)&=&{\rm Tr}\left\{{\bf\Sigma}_L^<(\epsilon){\bf
G}^>(\epsilon)-{\bf\Sigma}_L^>(\epsilon){\bf
G}^<(\epsilon)\right\},
\end{eqnarray}
where Hartree and Fock parts of self-energy components are
\begin{eqnarray}
{\bf\Sigma}^{H,r}&=&i\sum_\lambda\frac{2}{\Omega_\lambda}\int\frac{d\epsilon'}{2\pi}{\bf
M}^\lambda{\rm Tr}[{\bf G}^<(\epsilon'){\bf M}^\lambda]\label{SigmaHr}\\
{\bf\Sigma}^{H,<}&=&0\\
{\bf\Sigma}^{F,r}(\epsilon)&=&i\sum_\lambda\int\frac{d\epsilon'}{2\pi}{\bf
M}^\lambda[D^r_0(\epsilon-\epsilon'){\bf
G}^<(\epsilon')\nonumber\\
&\quad&+D^r_0(\epsilon-\epsilon'){\bf
G}^r(\epsilon')+D^<_0(\epsilon-\epsilon'){\bf G}^r(\epsilon')]{\bf M}^\lambda\\
{\bf\Sigma}^{F,<}(\epsilon)&=&i\sum_\lambda\int\frac{d\epsilon'}{2\pi}{\bf
M}^\lambda D^<_0(\omega-\omega'){\bf G}^<(\epsilon'){\bf
M}^\lambda.
\end{eqnarray}
Here the vibrational modes are labeled by $\lambda$, and
$\Omega_\lambda$ is the corresponding eigenfrequency. It is worth
noting that the lack of translational invariance makes the
retarded Hartree term non-zero, and potentially important.  Also,
at this stage the phonon propagators are undamped -- an
approximation that merits further investigation. The coupled
equations are iterated until convergence is achieved, and in the
following we give some representative results.
\begin{figure}\label{fig:twochains}\begin{center}
\includegraphics[width=8cm]{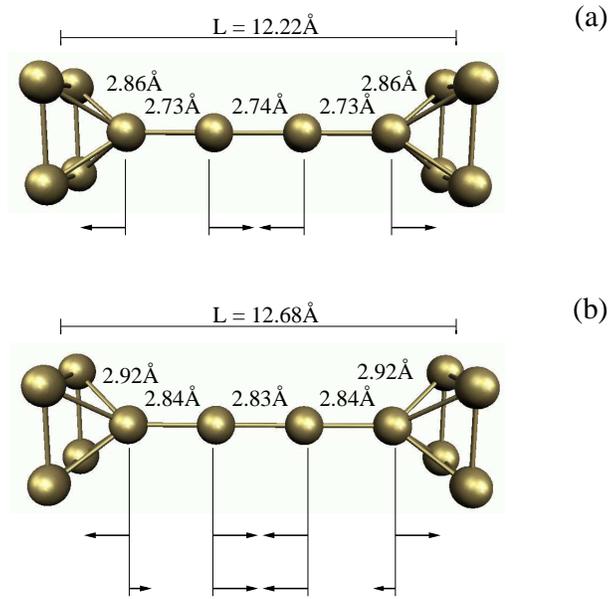}
\caption{Geometry of a
four-atom gold wire under two different states of stress.  The
dominating alternating bond-length modes, which cause the
inelastic scattering, are shown schematically with arrows. For the
shorter wire only one mode is active, while the elongated wire has
two active modes. (Reproduced from Ref.\protect\cite{Fred04}).}
\end{center}
\end{figure}
Let us consider a linear four-atom gold wire under two different
states of strain, as shown in Fig. 1. We calculate the phonon
signal in the nonlinear differential conductance vs. bias voltage
for two extremal cases: the energy transferred from the electrons
to the vibrations is either (i) instantaneously absorbed into an
external heat bath, or (ii) accumulated and only allowed to leak
via electron-hole pair excitations.  These limits are referred to
as the externally damped and externally undamped cases,
respectively. The next two figures show the corresponding results.
\begin{figure}[t]\label{fig:allmodes}\begin{center}
\epsfig{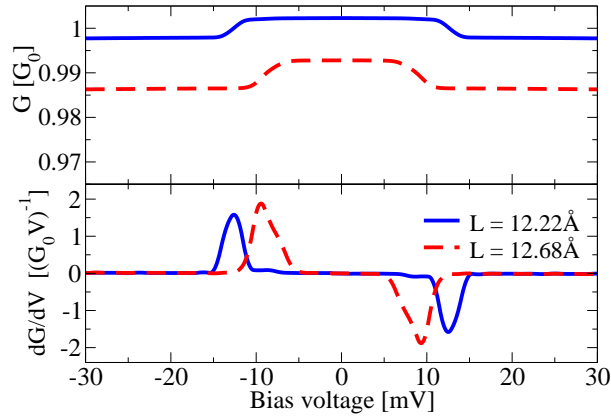} \caption{Differential
conductance and its derivative for the four-atom gold wire at two
different tensions in the case where the oscillators are
externally damped. Reproduced from Ref.\protect\cite{Fred04}.}
\end{center}\end{figure}
Since a typical
experiment is done at low temperatures, the mode occupation in the
externally damped case vanishes, $N_\lambda\approx 0$. In the
externally undamped case the mode occupation $N_\lambda$ is an
unknown parameter entering the electron-phonon self-energy,  and
additional physical input is necessary to determine this
parameter. We argue as follows. Since the system is in a steady
state, the net power transferred from the electrons to the device
must vanish, i.e., $P_L+P_R=0$. Using Eq.(\ref{PL}) one then
obtains the required  constraint on $N_\lambda$. This procedure
works in a straightforward way if there is only a single active
mode, but if several modes are present, a more detailed theory of
how the phonon modes equilibriate would be needed. The bottom
panel of Fig. 3 shows how the occupation of the leading phonon
mode, or, equivalently, the effective temperature, changes as a
function of bias.
\begin{figure}[t]\label{fig:onemode}\begin{center}
\epsfig{file=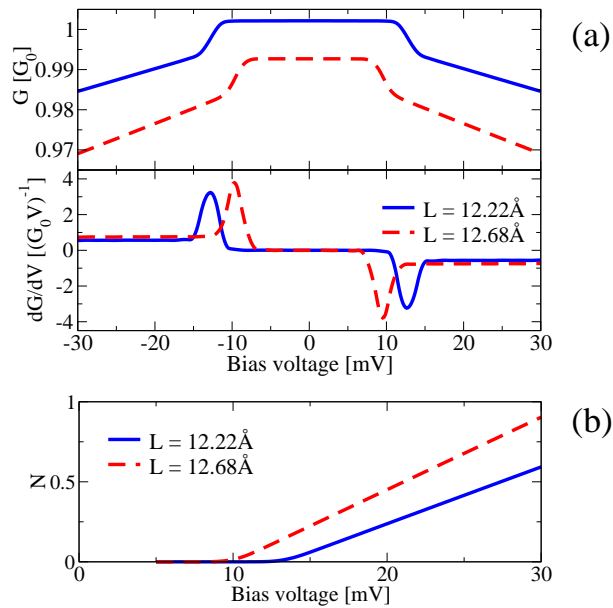,width=8cm} \caption{(a) Differental
conductance and its derivative for the four-atom gold wire at two
different tensions in the externally undamped limit.  Only the
most important mode is included in this calculation. (b) Mode
occupation vs. bias voltage. Reproduced from
Ref.\protect\cite{Fred04}.}\end{center}
\end{figure}
When comparing to the experiments of Agra{\"i}t et
al.\cite{Agra02}, one sees that the externally undamped model is
in near quantitative agreement with the data: the conductance drop
at the onset of inelastic scattering, and the slope after the drop
are very well reproduced. We view this as strong evidence of the
presence of heating in the experiment, but at the same time
recognize the need for a detailed microscopic theory including
phonon-phonon interactions.

\section{Conclusions} I have reviewed some of the
post-2002 developments in applying NEGF to modelling of transport
in mesoscopic systems. The common theme in my review has been to
focus on ``real devices", which may have ``real applications".  I
find it very pleasing that the NEGF technique, often regarded as
an academic exercise most suited for theoretical games, is now
becoming a strong tool in the analysis of practical devices, even
in industrial context.  At the same time there is still much room
for theoretical refinements, and I'm convinced that in the coming
years we will witness significant progress in this field, both
abstract and practical.

\section*{Acknowledgments}
The author  acknowledges fruitful collaborations with Mads
Brandbyge, Thomas Frederiksen, and Nicol{\'a}s Lorente, which led
to the results presented in Sec. \ref{sec:inelas}.  Thomas
Frederiksen made also a number of useful comments on the present
manuscript, for which the author is grateful.


\printindex
\end{document}